\documentclass[floatfix,%
 aip,
jmp,
 amsmath,amssymb,
 reprint,%
]{revtex4-1}
\usepackage[compatibility=false]{caption}
\usepackage{graphicx}
\usepackage{dcolumn}
\usepackage{natbib}
\usepackage{bm}
\usepackage{placeins}
\usepackage{subcaption}
\usepackage{graphicx}
\usepackage{xcolor}
\usepackage[compatibility=false]{caption}
\usepackage[version=4]{mhchem}

\usepackage[utf8]{inputenc}
\usepackage[T1]{fontenc}
\usepackage{mathptmx}

\begin{document}

\preprint{AIP/JAP}

\title[]{Exciton switching and tunable exciton-phonon coupling in Oxygen doped ZnO nanorods}
\author{Aiswarya Rath}
\author{Kalyan Ghosh}%
\affiliation{ 
School of Physical Sciences, National Institute of Science Education and Research (NISER) Bhubaneswar, Jatni, 752050, Odisha, India.
}%
\affiliation{
Homi Bhabha National Institute, Training School Complex, Anushakti Nagar, Mumbai 400094, India.
}%

\author{Pratap K Sahoo}
\email{pratap.sahoo@niser.ac.in}
\affiliation{ 
School of Physical Sciences, National Institute of Science Education and Research (NISER) Bhubaneswar, Jatni, 752050, Odisha, India.
}%
\affiliation{
Homi Bhabha National Institute, Training School Complex, Anushakti Nagar, Mumbai 400094, India.
}%
\affiliation{
Center for Interdisciplinary Sciences (CIS), NISER Bhubaneswar, Jatni-752050, Odisha, India.
}%

\date{\today}

\begin{abstract}
Self-ion doping in ZnO nanorods offers a powerful route to dynamically control excitonic states and light–matter interactions. Here, vertically aligned hexagonal ZnO nanorods were irradiated with 250 keV $O^{+}$ ions to establish a direct link between irradiation-induced defects and excitonic and optical responses. Increasing ion fluence progressively quenches the near-band-edge emission near 380 nm, leading to a complete exciton OFF state, while thermal annealing restores the emission, yielding a reversible exciton OFF/ON response. This switching originates from defect-mediated modification and thermal recovery of excitonic states. The defect-band emission (400–750 nm) exhibits quasi-periodic whispering-gallery-mode-like resonances that are progressively suppressed as irradiation alters nanorod geometry and optical confinement. Low-temperature photoluminescence resolves $D^{0}X$, FX–1LO, and FX–2LO transitions, whose temperature evolution follows the Bose–Einstein model. The non-monotonic variation of excition-phonon coupling strength reveals a competition among irradiation-induced defects, lattice distortion, and exciton localization in determining exciton–phonon coupling. These results establish $O^{+}$-ion irradiation as a means to simultaneously engineer excitonic switching, exciton–phonon interactions, and optical resonances, providing a pathway toward defect-programmable ZnO nanophotonic and optoelectronic devices. 
\end{abstract}


\maketitle

\section{Introduction} 
Zinc oxide (ZnO) is a wide-bandgap semiconductor of considerable interest for studying excitonic and electron–phonon phenomena, due to its direct bandgap of approximately 3.37 eV and large exciton binding energy of about 60 meV \cite{zgr2005, Janotti2009}. These characteristics enable excitonic states to remain stable at room temperature, making ZnO nanostructures an attractive platform for investigating exciton dynamics, exciton–phonon coupling, and the interplay between intrinsic electronic states and defect-induced states. The optical characteristics of ZnO are strongly influenced by excitonic interactions and can be tuned by reducing the dimensionality, manipulating native defects, and introducing foreign species into the lattice, all of which alter the electronic structure and the strength of exciton–phonon coupling (EPC) \cite{Klingshirn2007, Tainoff2010}.EPC is a fundamental interaction in polar semiconductors such as ZnO, where strong electron–phonon coupling can give rise to pronounced LO phonon replicas of excitonic transitions and, under appropriate conditions, enable phonon-assisted emission and stimulated processes. Numerous studies have demonstrated that EPC strength is sensitive to dimensionality, defect density, and carrier concentration, with 1D ZnO nanostructures often exhibiting enhanced EPC relative to bulk due to quantum confinement and increased surface/defect inclusions. Mn-doped ZnO nanowires show pronounced exciton–LO phonon interactions and 2LO-phonon assisted applications\cite{Sikdar_2021, sikdar1}. The one-dimensional geometry and large surface-to-volume ratio of ZnO nanorods amplify the effects of lattice disorder and defect states on exciton relaxation and phonon-assisted optical processes. However, realizing dynamic control over excitonic states—such as switching between free and bound excitons and modulating EPC remains a critical challenge for advancing tunable photonic devices. Oxygen doping, which introduces defect states and alters lattice dynamics, offers a promising pathway to engineer these properties without compromising the advantages of the one-dimensional nanostructure. 

In intrinsic ZnO nanorods (NRs), excitons exhibit strong coupling with longitudinal optical (LO) phonons owing to the polar nature of the crystal lattice. This interaction gives rise to characteristic exciton--phonon sidebands and, in certain cases, Fano-like interference features in photoluminescence (PL) spectra resulting from the coupling between discrete excitonic transitions and the phonon continuum. Incorporation of excess oxygen perturbs this interaction by modifying the native defect landscape through the formation of oxygen interstitials ($O_i$), oxygen-related defect complexes, or changes in zinc-vacancy concentrations. These defect states locally distort the lattice, alter the vibrational spectrum, and consequently modify exciton--phonon coupling. Unlike transition-metal doping, which introduces additional electronic states associated with impurity ions, oxygen incorporation uniquely leverages intrinsic lattice oxygen to achieve reversible exciton switching—a phenomenon where excitons toggle between localized defect-trapped states and delocalized free states under varying excitation conditions. Such defect-mediated excitonic reconfiguration provides an effective pathway for tuning the optical response of ZnO nanostructures while preserving their wide-bandgap semiconducting characteristics.

The exceptionally large exciton binding energy of ZnO, substantially exceeding the thermal energy at room temperature (RT), stabilizes excitonic states under ambient conditions and provides a favorable platform for probing exciton-related phenomena. Consequently, ZnO NRs exhibit efficient room-temperature excitonic emission, making them attractive for ultraviolet (UV) and blue optoelectronic devices, particularly low-threshold excitonic lasers. Depending on the geometry of the gain medium and the optical feedback mechanism, ZnO-based lasers are generally classified into three categories: random lasers, Fabry--Perot (FP) lasers, and whispering-gallery-mode (WGM) lasers. The ZnO nanopowders are used for Random lasing, where the nanoparticles simultaneously serve as the gain medium and multiple light scatterers, producing coherent feedback through recurrent scattering events within the disordered structure \cite{Padiyakkuth2022, Hsu2005}. In contrast, Fabry--Perot lasing is achieved in well-defined one-dimensional or planar ZnO architectures, including microbelts \cite{Zapien2004}, thin films \cite{Ni2015}, nanocombs \cite{Chen2011}, and related micro/nanostructures, where reflections from parallel end facets provide optical feedback.WGM resonators employ highly symmetric hexagonal ZnO micro- and nanostructures, such as microrods \cite{Yang2019} and hexagonal microdisks \cite{Gargas2010}, in which light is confined by repeated total internal reflection along the cavity perimeter. Owing to their exceptionally high quality factors, WGM resonators exhibit superior performance, including lower lasing thresholds, enhanced emission intensity, improved beam directionality, and narrower spectral linewidths, making them particularly attractive for high-performance ZnO-based photonic devices. On the other hand the reversible exciton OFF/ON behavior could potentially be exploited for optical memory, reconfigurable photonic circuits, and UV information encoding, provided the switching speed, cycling stability, and spatial addressability are established.

This study focused on the exciton switching effect and modulation of excitons by introducing defects through oxygen ion implantation on hydrothermally grown ZnO NRs. We examine how oxygen incorporation modifies the defect environment in ZnO NRs and show that these oxygen-related defects can induce transitions between excitonic states and enable systematic control of the EPC strength. Experimentally, we have observed that surface of the NRs modified from hexagonal to circular after post-implantation annealing, which affects excitation evolution as well as WGM variations. We propose that the excitations can be turned on and off as needed through thermal stimuli, serving as an exciton switch. Low-temperature PL spectroscopy was employed to probe the excitonic and defect-related states in the NRs and to realize the evolution of excitonic states in the presence of controllable scattering centers. 
\section{Experimental Techniques} 
Vertically aligned ZnO nanorods (NRs) were synthesized on (100)-oriented n-type Si substrates using a conventional hydrothermal method, following previously reported procedures \cite{Sikdar_2021, Singh1}. Zinc nitrate hexahydrate and hexamethylenetetramine (HMTA) were used as the precursor materials. Prior to the hydrothermal growth, the Si substrates were coated with a ZnO seed layer of approximately 30 nm thickness to facilitate uniform and preferentially oriented growth. The nanorods were subsequently grown hydrothermally for 6 h. The as-grown ZnO NRs were irradiated with 250 keV O$^{+}$ ions at five different fluences spanning 5×10$^{14}$ to 5×10$^{16}$ ions/cm$^{2}$. Following ion implantation, the samples were thermally annealed at 500$^\circ$C for 1 h under vacuum, with a base pressure of 5×10$^{-4}$ mbar. The surface morphology and structural characteristics of the NRs were examined before and after ion irradiation and post-irradiation annealing. Surface morphology was investigated using a field-emission scanning electron microscope (FESEM; Zeiss SIGMA), while the crystal structure was analyzed using a Rigaku X-ray diffractometer equipped with Cu K-$\alpha$ radiation ($\lambda$= 1.5418 Å). The optical properties and excitonic behavior of the ZnO NRs were investigated by PL spectroscopy. PL measurements were performed using a 325 nm He–Cd laser in a backscattering configuration, with a 40$\times$ UV objective. The emitted radiation was dispersed using a 1200 lines mm$^{-1}$ grating and detected with a CMOS-CCD detector. Temperature-dependent PL measurements were carried out using a Linkam THMS 600 cooling stage. At each measurement temperature, the sample was allowed to reach thermal equilibrium for 2 min before acquiring the spectrum. Monte Carlo simulations were performed using SRIM-2013 to determine the implantation characteristics and depth distribution of 250 keV $O^{+}$ ions in ZnO. The associated radiation-induced atomic displacements and displacements per atom (dpa) were evaluated using the SRIM/TRIM framework \cite{srim}. In addition, the fluence-dependent evolution of the target surface, including sputtering yields and compositional changes induced by ion bombardment, was simulated using the dynamic TRIDYN code \cite{tridyn}.

\section{Results and discussion}
\subsection{Ion Beam Simulations}
The depth-dependent distribution of implanted O$^+$ ions in the ZnO NRs was evaluated using SRIM/TRIM simulations in the Full Cascade mode, as presented in Fig. 1. The calculations yield a projected range ($R_P$) of approximately 414 nm for 250 keV O$^+$ ions. The calculated ratio of electronic to nuclear stopping power ($S_e/S_n$), projected range, and longitudinal straggling are 0.176, 102.8 nm, and 29.9 nm, respectively. The simulated ion profile therefore indicates an oxygen-enriched region extending to a depth of approximately 580 nm, with a longitudinal spread characterized by a standard deviation of 29.9 nm. At an implantation energy of 250 keV, nuclear stopping is substantially stronger than electronic stopping, as indicated by the relatively small $S_e/S_n$ ratio. Consequently, a significant fraction of the incident ion energy is deposited through collisions with the ZnO lattice, producing atomic displacements and thereby generating Zn and O vacancies and interstitials. Figure 1(a) presents the calculated depth profiles of the implanted O ions and the corresponding displacements per atom (dpa). The defect production rate increases with depth and reaches a maximum at approximately 350 nm, followed by a gradual decrease beyond this depth, and becomes comparatively small near 550 nm. The defect distribution was obtained using the SRIM/TRIM framework, which employs the Kinchin–Pease approach for estimating displacement damage \cite{srim_dpa1}. The dpa was calculated from the implanted-ion fluence and the number of vacancies generated per incident ion according to the reference \cite{raman_Zn_defect1}:
The atomic density of the target (ZnO) material used for dpa calculation was $8.30 \times 10^{22}$ atoms/cm$^3$. 

\begin{figure}[!ht]
\centering
\includegraphics[width=1\linewidth]{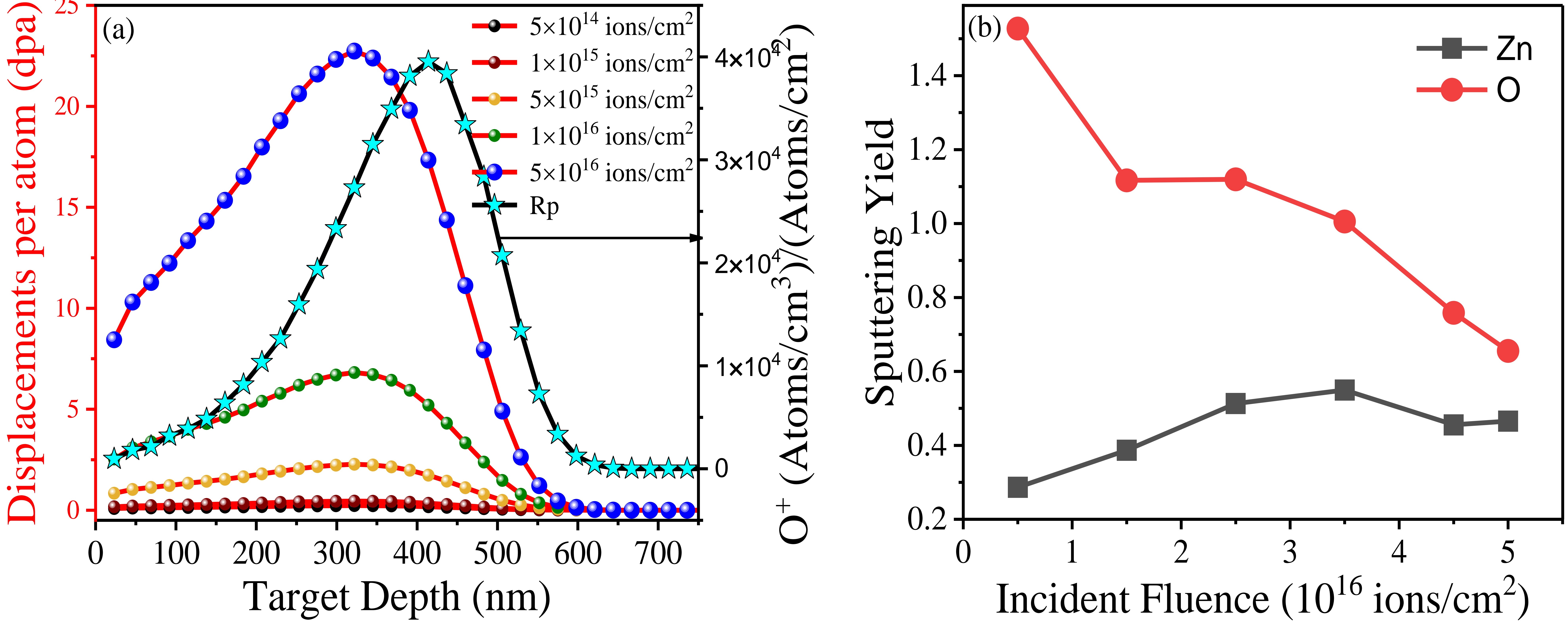}
\caption{(a) SRIM simulation showing penetration of 250 keV O$^{+}$ ions along the depth of ZnO NRs in right side of the y-axis and the left side shows dpa at different ion fluence, (b) Sputtering Yield data after O$^{+}$ ion implantation in ZnO NRs.}
\label{Theoretical simulations}
\end{figure}

The fluence-dependent partial sputtering yields of O and Zn, obtained from TRIDYN simulations, are shown in Fig. \ref{Theoretical simulations}(b). At the maximum implantation fluence of $\Phi = 5\times10^{16}$ ions/cm$^2$, the calculated sputtering removes approximately 50 nm from the ZnO NRs surface. At the initial stage of irradiation, the sputtering yield of O is considerably higher than that of Zn. The incident O$^{+}$ ions can efficiently transfer energy to lattice O atoms because they have the same mass, which may contribute to preferential oxygen removal. Although the O sputtering yield decreases from 1.5 to 0.6 atoms/ion while the Zn yield increases from 0.3 to 0.5 atoms/ion, oxygen removal remains greater, potentially favouring oxygen-vacancy formation. With increasing fluence, however, the surface composition progressively evolves toward a steady-state condition, resulting in a convergence of the O and Zn sputtering yields \cite{tridyn}. The preferential sputtering of oxygen modifies the near-surface stoichiometry, reducing the O atomic fraction and leading to a relative enrichment of Zn. Consequently, the irradiated surface region becomes increasingly oxygen-deficient as ion fluence increases, thereby rendering it partially amorphous \cite{tridyn}. In contrast, the implanted O$^{+}$ ions produce an oxygen-enriched region extending to approximately 532 nm. This spatially nonuniform oxygen distribution is expected to modify the electronic character of the nanorods: the oxygen-rich region can exhibit p-type behavior, whereas the underlying region retains n-type character associated with donor-like intrinsic defects, particularly oxygen vacancies ($V_{o}$) and zinc-interstitials ($Zn_{i}$) \cite{PhysRevB.76.165202}. Thus, ion implantation produces a depth-dependent modulation of the chemical composition and defect landscape of the ZnO nanorods, giving rise to a p–n junction-like architecture in which the oxygen-rich, relatively unmodified regions constitute the p-type and n-type components, respectively.

\subsection{Structural properties of vertically aligned ZnO Nanorods}

\begin{figure}[!hb]
\centering
\includegraphics[width=1\linewidth]{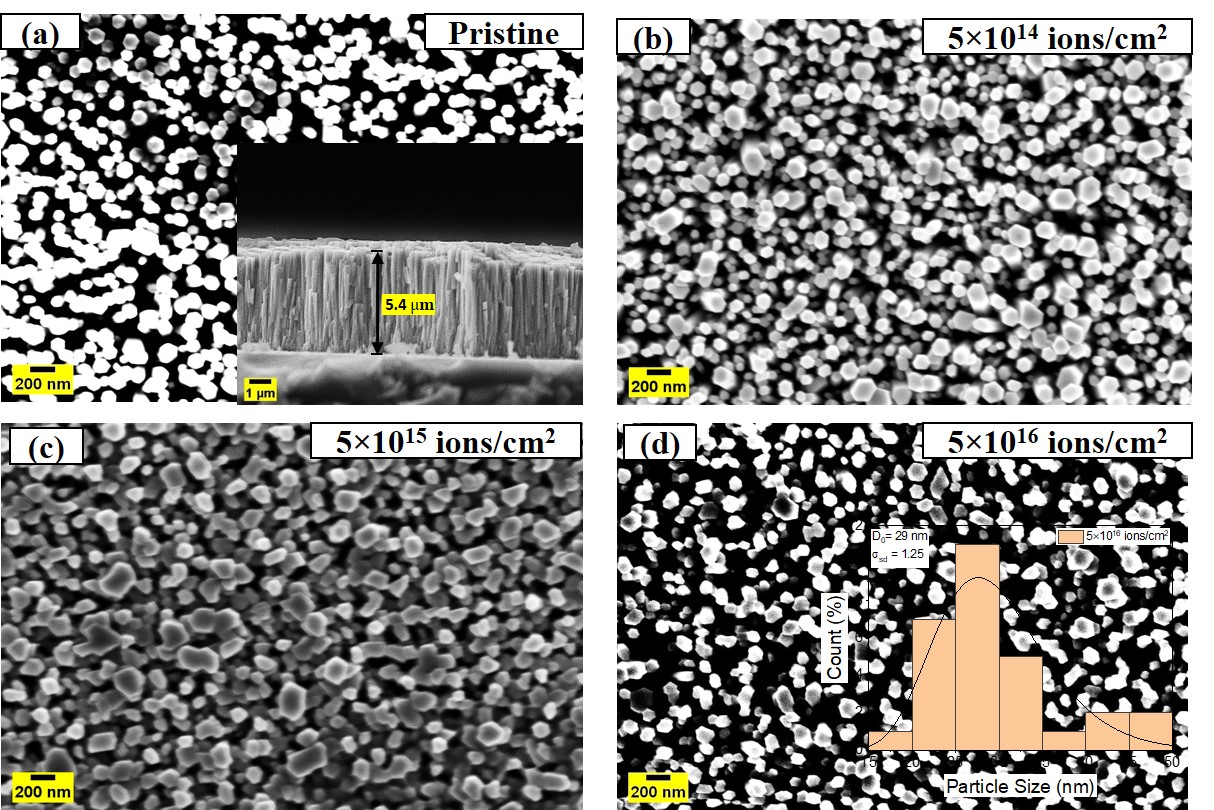}
\caption{FESEM images of the (a) pristine and O$^{+}$ ion implanted vertically aligned ZnO Nanorods at ion fluences (b) $5 \times 10^{14}$ ions/cm$^2$ , (c) $5 \times 10^{15}$ ions/cm$^2$ and (d) $5 \times 10^{16}$ ions/cm$^2$ .}
\label{FESEM1}
\end{figure}

\begin{figure}[!ht]
\centering
\includegraphics[width=1\linewidth]{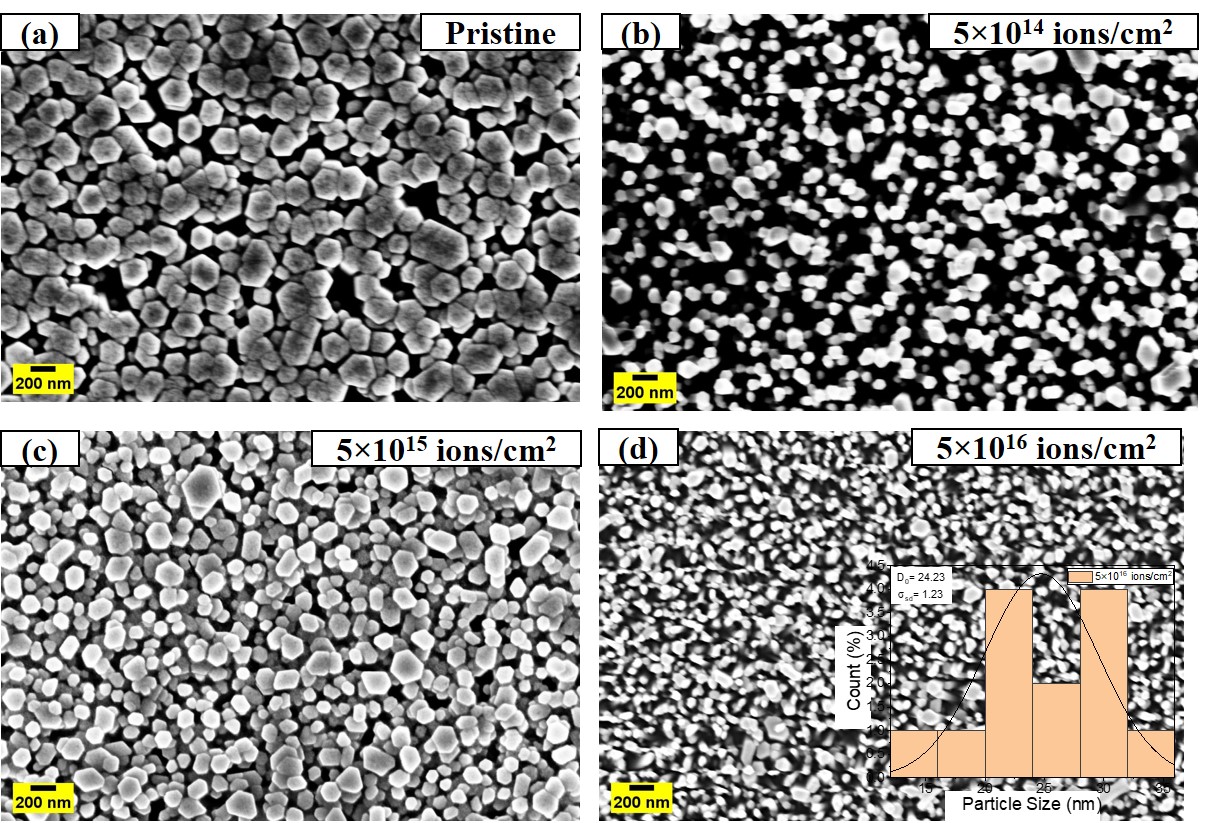 }
\caption{FESEM images showing the morphological evolution of ZnO nanorods following post-irradiation annealing at $500^\circ\text{C}$ (a) pristine and O$^{+}$ ion implanted ZnO Nanorods at ion fluences (b) $5 \times 10^{14}$ ions/cm$^2$ , (c) $5 \times 10^{15}$ ions/cm$^2$ and (d) $5 \times 10^{16}$ ions/cm$^2$.}
\label{FESEM2}
\end{figure}

Figure \ref{FESEM1} presents the top-view FESEM images of the vertically aligned ZnO NRs irradiated with different fluences of O$^{+}$ ions. The inset of Fig. \ref{FESEM1}(a) indicates a cross-sectional FESEM image with an average NR height of $5.4\pm1.2~\mu$m for the pristine sample. The initially well-defined hexagonal morphology of the NR tips progressively evolves with increasing ion fluence, as evident in Fig. \ref{FESEM1}(b)–(d). At the highest implantation fluence, the NR cross sections become predominantly circular or irregular rather than hexagonal. The corresponding diameter distribution shown in the inset of Fig. \ref{FESEM1}(d) reveals a reduction in the mean NR diameter from $42.12 \pm 1.25$ nm to $29.0 \pm 1.25$ nm with increasing ion fluence, indicating substantial modification of the near-surface morphology by ion irradiation.

\begin{figure}[!ht]
\centering
\includegraphics[width=0.9\linewidth]{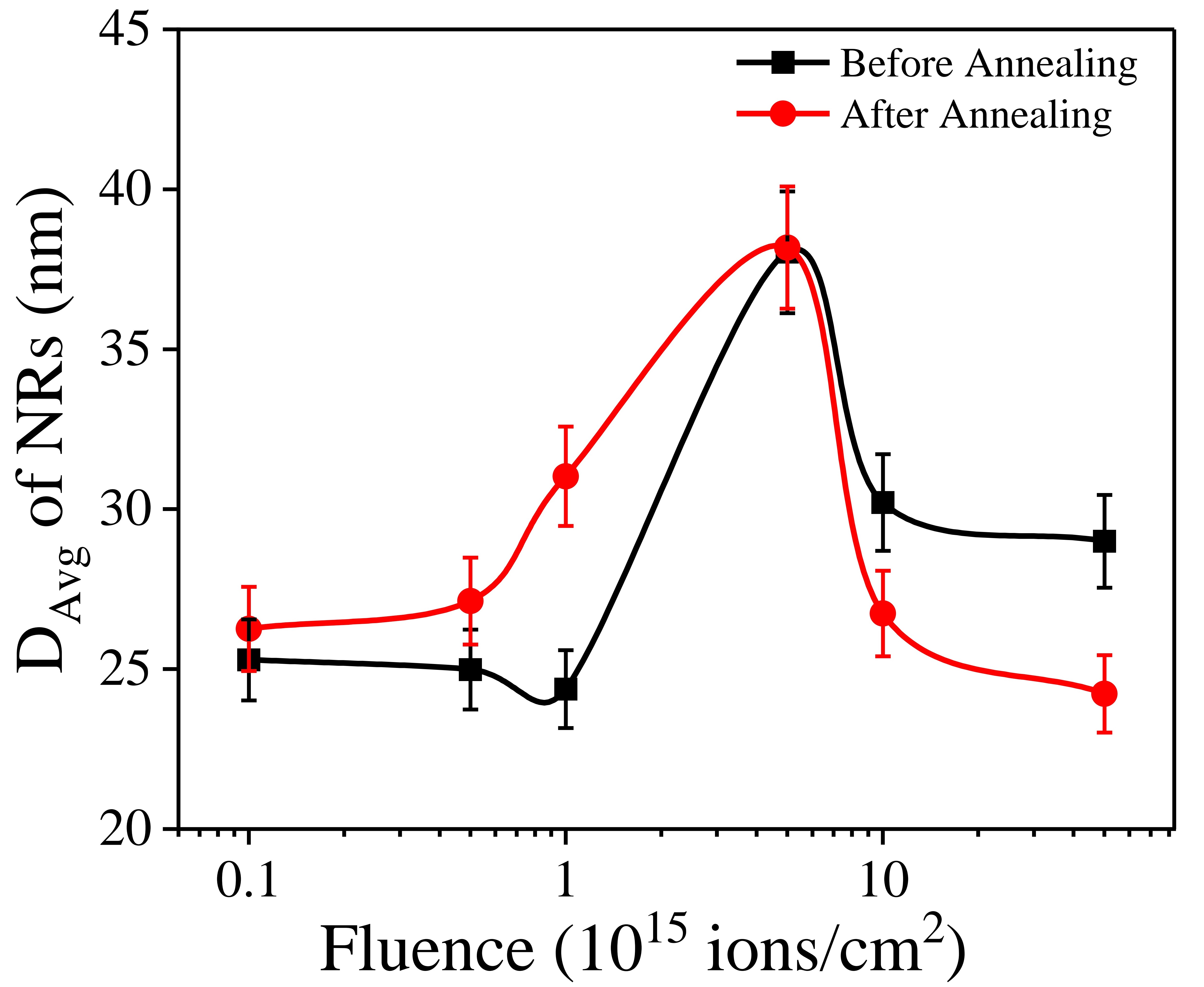}
\caption{The average diameter of the ZnO NRs after O$^{+}$ ion implantation at different fluences and post annealing.}
\label{size-dis}
\end{figure}

The effect of post-implantation thermal treatment is illustrated by the FESEM images in Fig. \ref{FESEM2}, which correspond to the samples annealed at $500^\circ$C. Thermal annealing further modifies the morphology of both pristine and irradiated NRs, resulting in a pronounced deviation from the initial hexagonal cross-sectional geometry. Following irradiation and annealing, the mean NR diameter decreases from $52.12 \pm 1.25$ nm to $24.23 \pm 1.23$ nm over the investigated fluence range. Interestingly, the average diameter $(D_{Avg})$ exhibits a non-monotonic evolution with ion fluence, reaching a maximum around $\Phi = 5 \times 10^{15}$ ions/cm$^{2}$, followed by a subsequent decrease at higher fluences as shown in Fig. \ref{size-dis}. This crossover suggests a competition between irradiation-induced surface modification and thermally activated processes during annealing, leading to a fluence-dependent evolution of the NR morphology.

\begin{figure}[!ht]
\centering
\includegraphics[width=0.9\linewidth]{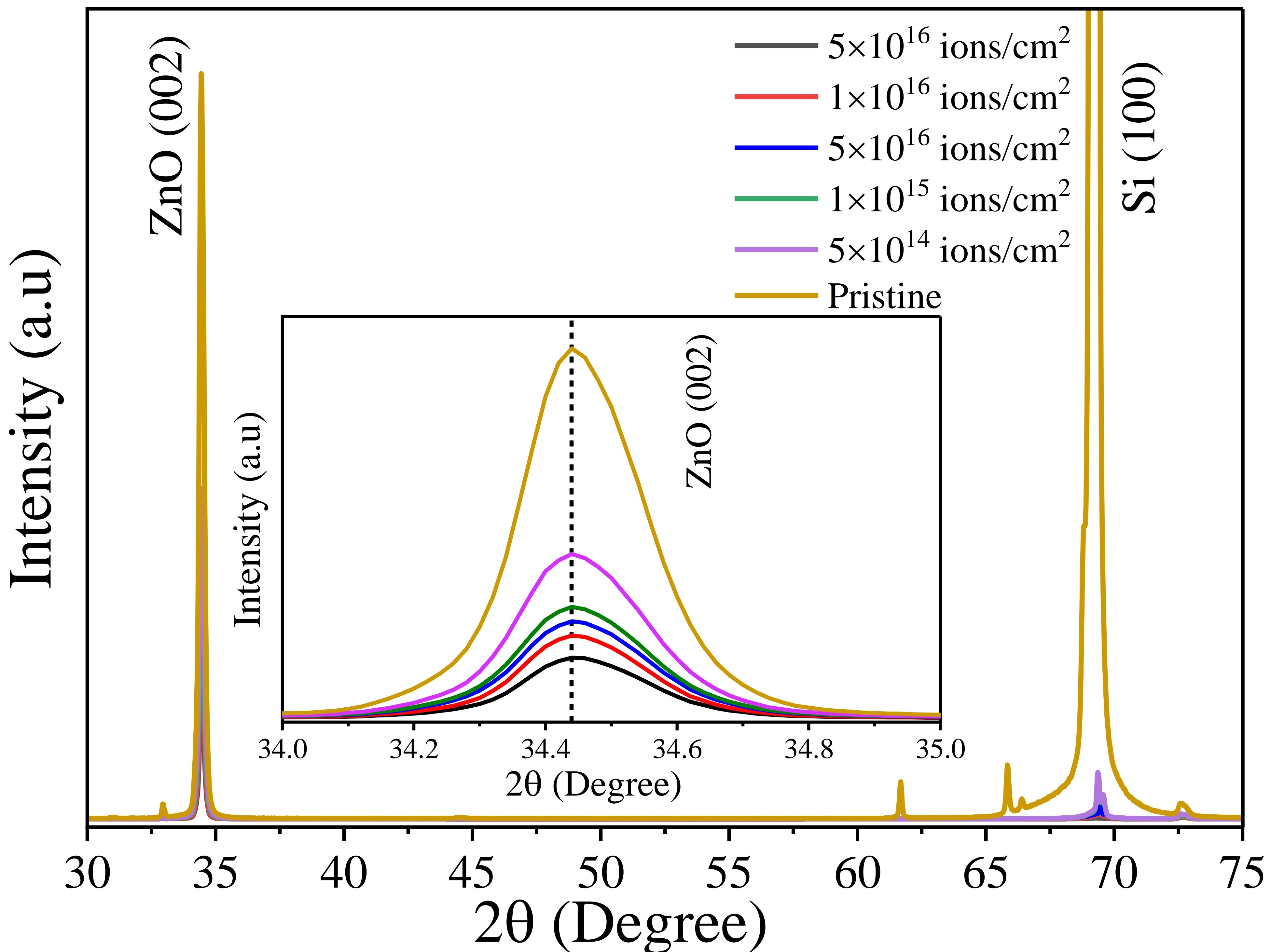}
\caption{XRD pattern of pristine and 250 keV O$^{+}$ ion implanted samples after annealing at $500^\circ$C.}
\label{XRD}
\end{figure}

Figure \ref{XRD} shows the X-ray diffraction patterns of the pristine and O$^+$-implanted ZnO NRs following annealing at $500^\circ$C. The pristine NRs exhibit a dominant diffraction peak at approximately $34.48^\circ$, which can be indexed to the (002) reflection of the hexagonal wurtzite ZnO structure. The pronounced intensity of this reflection confirms the preferential orientation of the nanorods along the crystallographic $c$-axis. The diffraction peak at approximately $69.24^\circ$ originates from the Si (100) substrate. Two relatively weak ZnO reflections at approximately $31.14^\circ$ and $62.68^\circ$, assigned to the (100) and (103) planes, respectively, are observed only for the pristine sample. Following O$^+$ implantation and subsequent annealing, these weaker reflections are no longer discernible, while the characteristic (002) reflection remains present for all implantation fluences. This indicates that the hexagonal wurtzite structure is retained after ion irradiation without detectable formation of a secondary crystalline phase. Nevertheless, the intensity of the (002) reflection decreases progressively with increasing ion fluence. This reduction can be correlated with the increasing radiation-induced displacement damage, as indicated by the SRIM simulations in Fig. \ref{Theoretical simulations}(a). The enhanced dpa at higher fluences introduces a greater concentration of lattice defects and structural disorder within the near-surface region of the nanorods. The simulated implantation profile further indicates that the implanted oxygen is concentrated predominantly within the upper few hundred nanometres of the NRs. The fluence-dependent suppression of the (002) diffraction intensity therefore suggests increasing structural disorder and partial loss of long-range crystalline order in the near-surface region. Although the wurtzite framework remains identifiable by the persistence of the (002) reflection, the reduced diffraction intensity indicates progressive degradation of the crystallographic order as O$^+$ fluence increases. The post-implantation annealing does not completely restore the original diffraction intensity, suggesting that a fraction of the irradiation-induced disorder persists after thermal treatment.

The crystallite size ($D$), microstrain ($\epsilon$), and dislocation density ($\delta$) were estimated from the XRD profiles, and the corresponding values are summarized in Table \ref{Table}. The pristine ZnO NRs exhibit a crystallite size of 41.65 nm, which decreases to approximately 38 nm following O$^+$ implantation at fluences up to $1\times10^{16}$ ions/cm$^{2}$. At the highest fluence of $5\times10^{16}$ ions/cm$^{2}$, the crystallite size decreases substantially to 26.97 nm, indicating enhanced disruption of the crystalline domains under intense ion irradiation. The microstrain increases from $8.689\times10^{-4}$ for the pristine NRs to a maximum value of $9.534\times10^{-4}$ at a fluence of $1\times10^{15}$ ions/cm$^{2}$. With further increases in fluence, the microstrain decreases modestly, although it remains higher than that of the pristine sample. A similar evolution is observed for the dislocation density, which increases from $5.8\times10^{-4}$ nm$^{-2}$ in the pristine sample to $6.9\times10^{-4}$ nm$^{-2}$ at $5\times10^{14}$ and $1\times10^{15}$ ions/cm$^{2}$. At higher fluences, $\delta$ gradually decreases, reaching $6.2\times10^{-4}$ nm$^{-2}$ at $5\times10^{16}$ ions/cm$^{2}$. The initial increase in microstrain and dislocation density can be associated with the accumulation of irradiation-induced lattice defects and atomic displacements. At higher fluences, the subsequent reduction in these parameters, together with the pronounced decrease in crystallite size, may reflect the development of substantial structural disorder and defect-rich regions within the irradiated nanorods. Overall, the XRD-derived parameters demonstrate that O$^+$ implantation progressively modifies the crystalline domain structure and introduces measurable lattice distortion in the ZnO NRs.

\begin{table}[!t]
\centering
\caption{Table showing the variation of Crystallite size (D), Microstrain ($\epsilon$) and Dislocation Density ($\delta$) with ion fluence estimated from the XRD pattern.}
\label{table XRD parameters}
\begin{tabular}{ccccccc}
\centering
Fluence$(ions/cm^{2})$&$2\theta\ (^\circ)$&$\beta$$(^\circ)$&D$(nm)$&$\epsilon(10^{-4})$&$\delta$$(10^{-4}nm^{-2})$\\
\hline
Pristine(0) &\mbox{34.44}&\mbox{0.2085}&\mbox{41.65}&\mbox{8.689}&\mbox{5.8}\\
$5\times10^{14}$&\mbox{34.44}&\mbox{0.2281}&\mbox{38.07}&\mbox{9.506}&\mbox{6.9}\\
$1\times10^{15}$&\mbox{34.44}&\mbox{0.2288}&\mbox{37.96}&\mbox{9.534}&\mbox{6.9}\\
$5\times10^{15}$&\mbox{34.44}&\mbox{0.2209}&\mbox{39.31}&\mbox{9.206}&\mbox{6.5}\\
$1\times10^{16}$&\mbox{34.44}&\mbox{0.2274}&\mbox{38.18}&\mbox{9.478}&\mbox{6.4}\\
$5\times10^{16}$&\mbox{34.44}&\mbox{0.3220}&\mbox{26.97}&\mbox{9.342}&\mbox{6.2}\\
\end{tabular} 
\label{Table}
\end{table}

\subsection{\label{sec:level2} Photoluminescence Spectra of ZnO Nanorods} 

\begin{figure}[!ht]
\centering
\includegraphics[width=0.9\linewidth]{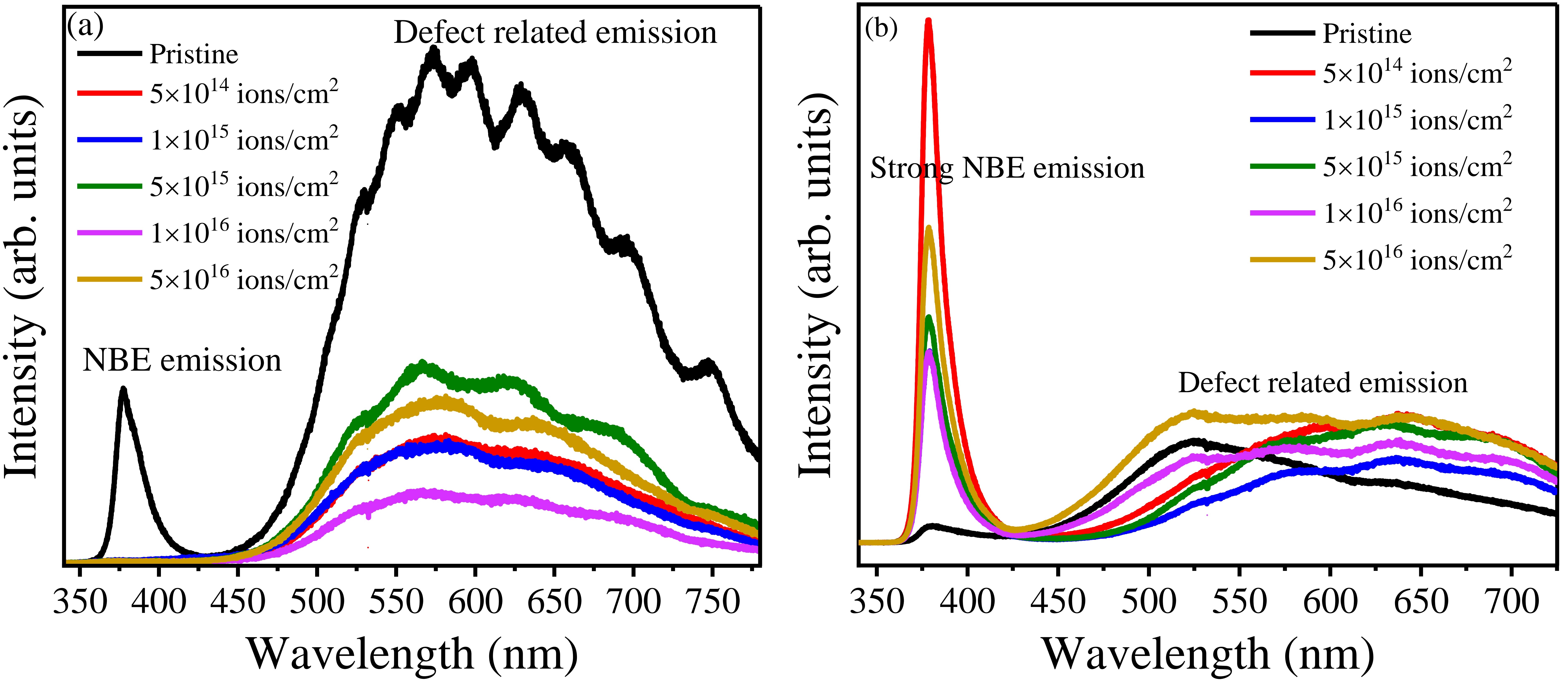}
\caption{Room temperature Photoluminescence Spectra of pristine and 250 keV O$^{+}$ ion implanted ZnO NRs (a) before and (b) after annealing.}
\label{PL RT}
\end{figure}

Photoluminescence spectroscopy is a non-destructive approach for probing excitonic emission and phonon replicas without significantly damaging the samples. Figure \ref{PL RT}(a) illustrates the room-temperature PL spectra of the ZnO NRs recorded under 325 nm excitation. Along with a prominent NBE emission, a strong defect-induced visible emission is also observed in the pristine sample, indicating that the crystalline ZnO NRs also contain optically active defect centers. In contrast, the NBE emission is completely suppressed in the O-ion-implanted samples, which can be ascribed to the introduction of implantation-induced defects that provide competing non-radiative recombination pathways and carrier trapping\cite{SINGH2017143}. This quenching of the NBE emission can therefore be regarded as an effective "switching-off" of the excitonic emission.

In addition to the NBE emission, all samples exhibit broad deep-level emissions in the visible spectral region, which are associated with intrinsic and implantation-induced defects, including zinc vacancies (V$_{Zn}$), oxygen vacancies (V$_O$), zinc interstitials (Zn$_i$), and oxygen interstitials (O$_i$). Interestingly, these defect peaks exhibit pronounced quasi-periodic, relatively narrow oscillations superimposed on the broad defect-related emission, which can be attributed to whispering-gallery-mode (WGM)-like resonances because of the hexagonal geometry of the ZnO NRs. The oscillatory PL features therefore provide evidence of the coupling between defect-related optical emission and the geometrically confined optical modes of the nanorods. Moreover, the distinct WGM-like oscillations gradually diminish with increasing ion fluence, suggesting progressive modification from the hexagonal shape of the nanorod optical cavity and enhanced defect-induced non-radiative losses.

Following thermal annealing of the implanted samples at 500$^\circ$C for 1 h, the NBE emission reappears with substantially enhanced intensity, indicating recovery of radiative excitonic recombination. This restoration of the NBE emission can consequently be interpreted as a "switching-on" of the excitonic emission by thermal energy through recovery of the implantation-induced defects. Consistent with the FESEM observations, the initially well-defined hexagonal top surface of the NRs becomes progressively rounded and approaches a circular morphology after post-implantation annealing. Consequently, the intensity of the emission in the range of 400-750 nm and the number of discernible whispering-gallery-mode (WGM) oscillation modes are significantly reduced, as shown in Fig. \ref{PL RT}(b). The reduction in the number and intensity of the oscillatory modes indicates that thermal energy modifies both the defect population and the optical confinement conditions within the NRs. The nearly periodic modulation of the emission envelope suggests the presence of WGM-like optical confinement. From the resonance positions observed in the range of approximately 550–750 nm, the characteristic free spectral range is estimated to be $\sim$ 30-50 nm, corresponding to a wavenumber spacing of approximately $0.8-1.2\times10^3 cm^{-1}$. Using the WGM relation \ $ \Delta \lambda \simeq \lambda ^ 2 / (2\pi n_{eff}R)$, with $n_{eff}\sim 2.0-2.1$ for ZnO in the visible spectral range, an effective resonator radius of approximately 0.7 to 0.9 $\mu$m is obtained. This corresponds to an estimated azimuthal mode number of $m\sim 15-20$ around 600 nm. The observed spectral modulation is strongly affected by $O^{+}$-ion irradiation, indicating that ion-induced defects modify the local dielectric environment, thus affecting the optical confinement and excitonic recombination pathways. The evolution of the WGM-like modulation together with the defect-related emission therefore provides evidence for defect-mediated modification of the photonic environment of the ZnO nanostructures.

\subsection{\label{sec:level2-temp} Temperature dependent study of NBE PL spectra} 

\begin{figure}[!htpb]
\centering
\includegraphics[width=0.9\linewidth]{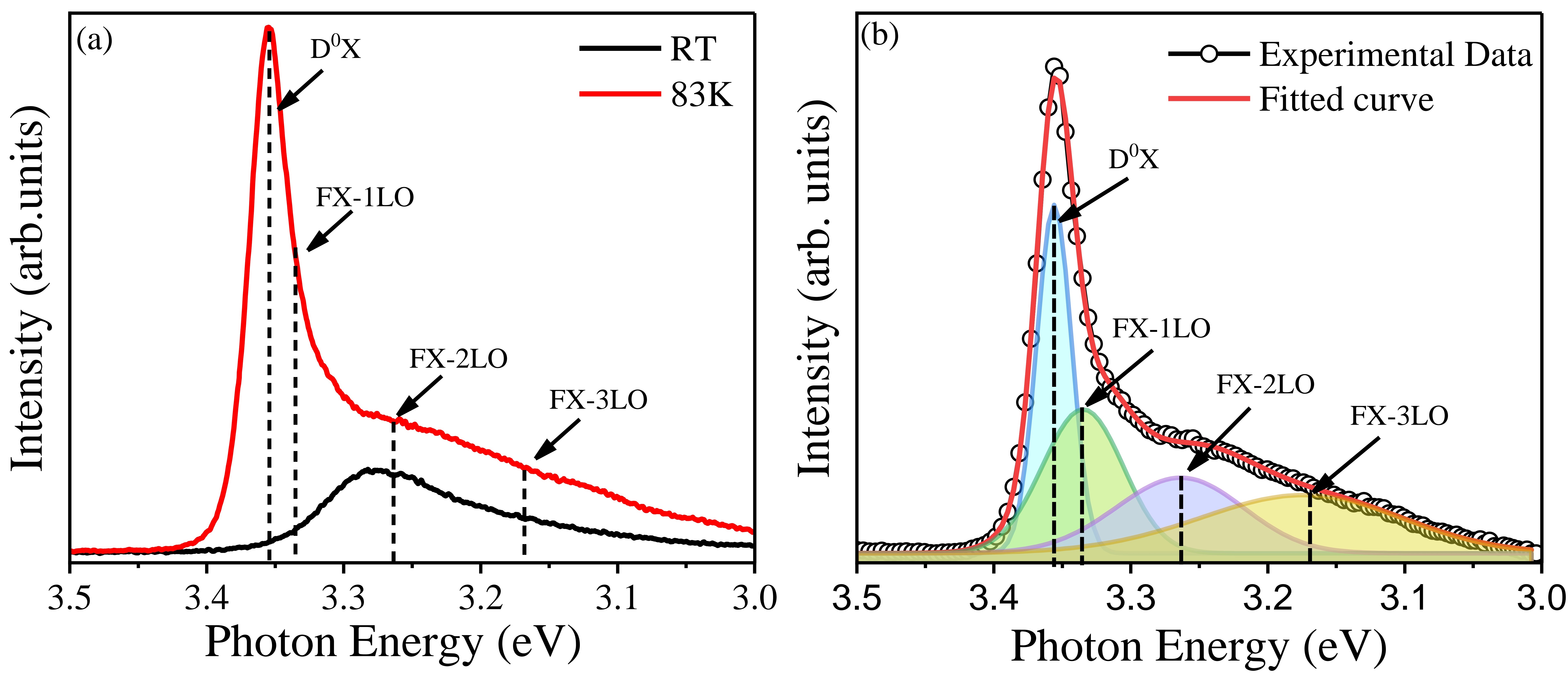}
\caption{(a) The comparison between NBE emission peak at room temperature and 83 K, respectively, and (b) the deconvoluted NBE-PL spectra at 83 K.}
\label{NBE peaks}
\end{figure}

\begin{figure*}[!ht]
\centering
\includegraphics[width=1\linewidth]{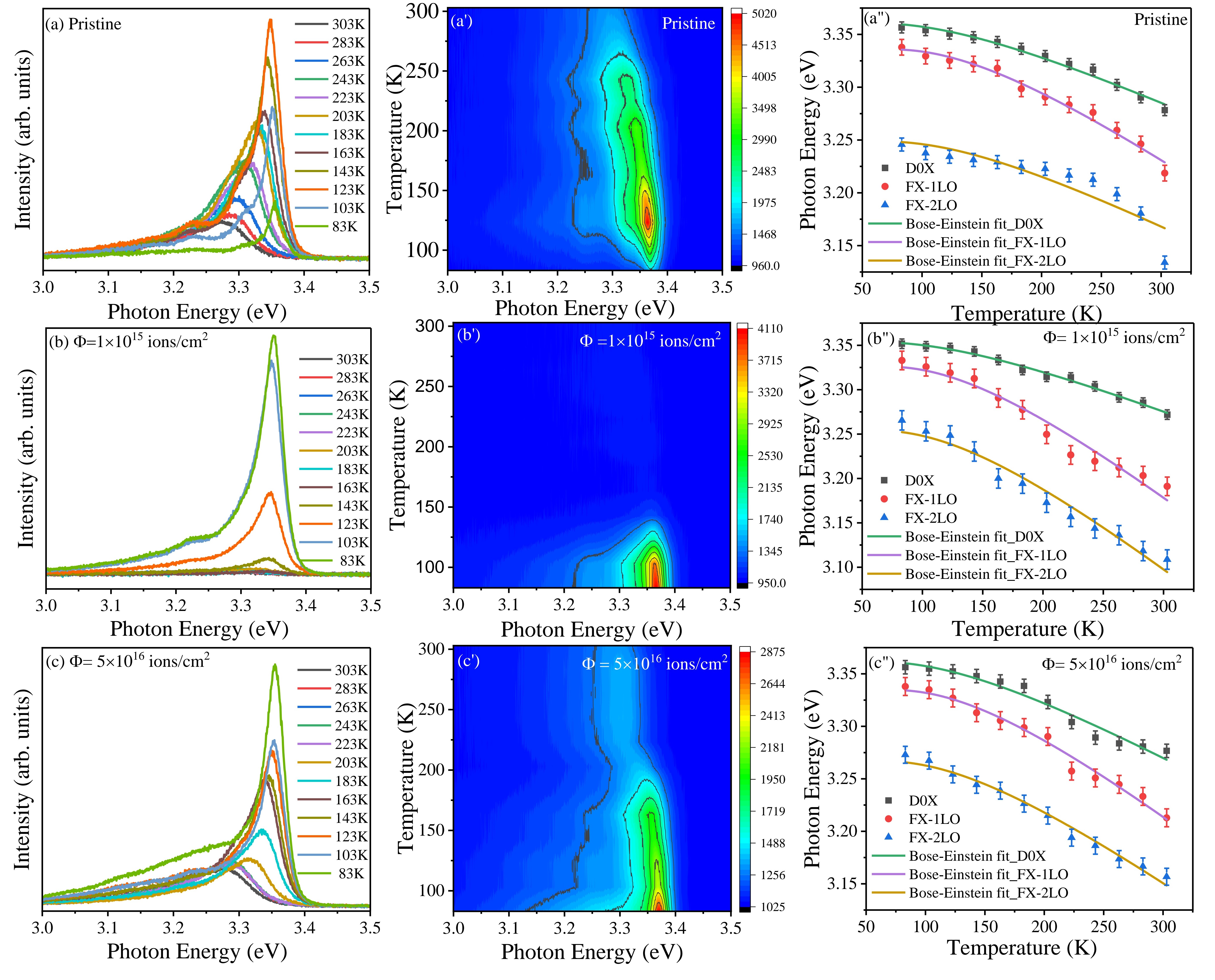}
\caption{Temperature dependent PL spectra of (a) Pristine (b) medium fluence @ $1\times10^{15}$ ions/cm$^{2}$ and (c) highest fluence @ $5\times10^{16}$ ions/cm$^{2}$ of ZnO NRs for the NBE emission region  ranging from 83 K to 303 K}
\label{Temp dep PL}
\end{figure*}

Although NBE emission is observed at RT, the peaks are not fully resolved due to stronger phonon interactions and scattering. To obtain distinct peaks, low-temperature PL measurements were performed using liquid Nitrogen to reach out of minimum temperature of 83K. The PL spectra are collected using a 325nm excitation Laser in the temperature range of 83-303K. This enables us to distinguish the excitonic peaks and the higher-order phonon replicas, FX-nLO. Figure \ref{NBE peaks}(a)  shows the NBE emission peak comparison between RT and at 83 K for the highest dose sample. It is clearly visible from the plot that the sharpness of peaks increases and can be deconvoluted into $D^oX$(donor-bound exciton) centered around 3.36 eV as well as 3 distinct phonon replicas, i.e., FX-nLO (n=1,2,3) sub-peaks as shown in Figure \ref{NBE peaks} (b).

The temperature-dependent PL spectra of the ZnO hexagonal NRs are presented column-wise in Fig. \ref{Temp dep PL}, for pristine and for $O^{+}$ ion irradiation fluences of $1 \times 10^{15}$ ions/cm$^2$, and $5 \times 10^{16}$ ions/cm$^2$. The left column (a–c) indicates PL spectra vs photon energy in the temperature range of 83-303K, the middle column ($a'$–$c'$) shows the temperature–energy intensity contour maps. The right column ($a''$–$c''$) illustrates the Peak energy evolution as a function of temperature using the Bose–Einstein fit. Temperature-dependent PL measurements were recorded to investigate the influence of $O^{+}$ ion irradiation on excitonic recombination dynamics of the vertically aligned ZnO NRs. The pristine sample shows a strong NBE emission centered at approximately 3.36 eV, which becomes sharper and more intense as the temperature decreases, as illustrated in Fig. \ref{Temp dep PL} ($a$--$c$) and ($a'$--$c'$), and explains the dominant nature of excitonic recombination with crystalline quality. At low temperatures, three emission components are substantially resolved to the neutral donor-bound exciton ($D^0X$) and the phonon-assisted free exciton replicas FX–1LO and FX–2LO. The progressive redshift of all exciton peaks with increasing temperature arises from band-edge shrinkage governed by exciton–phonon interaction. The extracted NBE peak energies are well described by the Bose–Einstein model, indicating a significant temperature-dependent renormalization of the ZnO band-edge transition arising from exciton–phonon interactions. This behavior is consistent with the high crystalline quality of the ZnO nanostructures\cite{bose_einstein_fit}.

After irradiation with a fluence of $1 \times 10^{15}$ $ions/cm^2$, a pronounced modification in the PL response has been observed. The emission intensity decreases significantly at higher temperatures, and the contour map reveals that radiative recombination is largely restricted to the low-temperature region ($<120$ K). This behavior indicates that irradiation-induced defects, such as oxygen vacancies and interstitial-related disorders, serve as non-radiative recombination centers. Consequently, free exciton emission is strongly thermally quenched, and enhanced exciton localization leads to a larger temperature-induced energy shift than in the pristine sample. The Bose–Einstein fitting indicates enhanced exciton–phonon coupling in the defect-engineered ZnO, which can be attributed to defect-induced lattice perturbations and the associated modification of the phonon environment.

At a higher fluence of $5 \times 10^{16}$ $ions/cm^2$, partial recovery of excitonic emission has been observed. Although spectral broadening persists, the NBE emission remains visible over a wider temperature range, implying defect restructuring via dynamic annealing induced by the thermal spike effect of 250 keV ion bombardment. The stabilization or clustering of irradiation-generated defects may reduce non-radiative recombination pathways, allowing localized excitonic recombination to dominate again. The temperature evolution of $D^0X$, FX–1LO, and FX–2LO transitions continues to obey Bose–Einstein behavior, confirming that exciton–phonon interaction remains the governing mechanism for bandgap renormalization. So we can say, $O^{+}$ ion irradiation induces a non-monotonic modification of optical properties in ZnO nanorods. Moderate fluence introduces significant lattice disorder and suppresses excitonic recombination, whereas higher fluence promotes defect reorganization, leading to partial restoration of radiative emission. 

At room temperature, the FX--2LO feature at approximately 3.36 eV is more intense than the $D^0X$ and FX--1LO features. This prominent phonon-assisted emission is consistent with strong exciton–LO-phonon coupling in polar ZnO. As temperature increases, thermal redistribution reduces the population of free excitons near the centre of the Brillouin zone, making direct radiative recombination less favourable relative to phonon-assisted recombination \cite{lo_phonon_energy2}. The temperature-dependent peak energies of $D^0X$, FX--1LO and FX--2LO are shown in figure~\ref{Temp dep PL}($a''$--$c''$). 

To understand the role of phonons for exciton recombination, the temperature dependence of $D^oX$ peak energies are fitted with a Bose--Einstein-type model \cite{bose_einstein_fit},

\begin{equation}\label{bose_einstein_equation}
E_x(T)=E_x(0)-
\frac{2a_{\mathrm{BE}}}{\exp(\Theta_{\mathrm{BE}}/T)-1},
\end{equation}

where $a_{\mathrm{BE}}$ characterizes the effective exciton--phonon interaction and $\Theta_{\mathrm{BE}}$ is a characteristic phonon temperature, represents the effective phonon energy scale through $k_{\mathrm B}\Theta_{\mathrm{BE}}$ and was fixed at 400 K for all fits, consistent with the value reported in Ref~\cite{Gu2007}, and $E_x(0)$ explains the excitonic peak position at absolute zero temperature. 

In the process of phonon-assisted exciton recombination, a part of the available energy is transferred to LO phonon modes, as observed in the experimental spectra as FX-1LO, FX-2LO, and FX-3LO. Since the FX-3LO intensity is very low, only the first two phonon replicas are used to fit the model hereafter. The complete Bose--Einstein expression is used for fitting the peak energies of $n$LO replica along with the $D^oX$ excitonic emission, given as follows\cite{bose_einstein_fit, lo_phonon_energy3}, 

\begin{equation}
\label{bose_einstein_replica}
\begin{aligned}
E_n(T) ={}&
\underbrace{
E_x(0)-\frac{2a_{\mathrm{BE}}}
{\exp(\Theta_{\mathrm{BE}}/T)-1}
}_{\text{temperature-dependent exciton energy}}
\\
&-\underbrace{
n\hbar\omega_{\mathrm{LO}}
}_{\text{energy transferred to LO phonons}}
\\
&+\underbrace{
\left(\frac{5}{2}-n\right)k_{\mathrm{B}}T
}_{\text{thermal shift of the replica maximum}}.
\end{aligned}
\end{equation}

where $\hbar \omega_{LO}$=72 meV represents the LO phonon energy of ZnO and the change in peak energy $\Delta E = \left( \frac{5}{2}-n \right)k_BT$. 
Free excitons have a distribution of centre-of-mass kinetic energies. Phonon-assisted recombination allows excitons with finite momentum to emit light, so the replica maximum depends on both the thermal exciton distribution and the energy-dependent transition probability. The measured and fitted replica peak reflects a fare competition between exciton-energy redshift and a thermal blueshift of the spectral maximum. 

\begin{table}[!ht]
\caption{Table depicting the extracted $E_x(0)$ and $a_{\mathrm{BE}}$ parameters from the Bose-Einstein type model for the free exciton and phonon recombination replica peaks from the PL spectra.}
\label{table PL extracted parameters}
\begin{tabular}{ccccccc}

Fluence $(ions/cm^2)$&Peak&$E_x(0)$(eV)&$a_{\mathrm{BE}}$(eV)\\
\hline
  &\mbox{$D^0X$}&\mbox{3.361}&\mbox{0.106}\\
Pristine(0) &\mbox{FX-1LO}&\mbox{3.400}&\mbox{0.190}\\
  &\mbox{FX-2LO}&\mbox{3.390}&\mbox{0.128}\\
  \hline
&\mbox{$D^0X$}&\mbox{3.333}&\mbox{0.109}\\
$5\times10^{14}$&\mbox{FX-1LO}&\mbox{3.352}&\mbox{0.242}\\
&\mbox{FX-2LO}&\mbox{3.374}&\mbox{0.201}\\
\hline
&\mbox{$D^0X$}&\mbox{3.354}&\mbox{0.110}\\
$1\times10^{15}$&\mbox{FX-1LO}&\mbox{3.391}&\mbox{0.250}\\
&\mbox{FX-2LO}&\mbox{3.396}&\mbox{0.234}\\
\hline
&\mbox{$D^0X$}&\mbox{3.355}&\mbox{0.094}\\
$5\times10^{15}$&\mbox{FX-1LO}&\mbox{3.414}&\mbox{0.225}\\
&\mbox{FX-2LO}&\mbox{3.401}&\mbox{0.154}\\
\hline
&\mbox{$D^0X$}&\mbox{3.354}&\mbox{0.091}\\
$1\times10^{16}$&\mbox{FX-1LO}&\mbox{3.393}&\mbox{0.200}\\
&\mbox{FX-2LO}&\mbox{3.395}&\mbox{0.150}\\
\hline
&\mbox{$D^0X$}&\mbox{3.347}&\mbox{0.082}\\
$5\times10^{16}$&\mbox{FX-1LO}&\mbox{3.393}&\mbox{0.194}\\
&\mbox{FX-2LO}&\mbox{3.397}&\mbox{0.141}\\

\end{tabular} 
\end{table}

The parameters extracted from the Bose--Einstein analysis are listed in table~\ref{table PL extracted parameters}. The uncertainties in $E_x(0)$ and $a_{\mathrm{BE}}$ are $\pm0.010$ and $\pm0.002$ eV, respectively. 
The characteristic phonon temperature, $\Theta_{\mathrm{BE}}$, was fixed at 400 K for all fits. The extrapolated $D^0X$ energy shifts from $3.361\pm0.010$ eV in pristine ZnO to $3.333\pm0.010$ eV at $5\times10^{14}$ ions/cm$^{2}$ and partially recovers at intermediate fluences. The smaller variations at higher fluences are comparable to the energy uncertainty. For the FX--1LO and FX--2LO fits, $E_x(0)$ represents the extrapolated energy of the parent free-exciton. 

The effective interaction parameter $a_{\mathrm{BE}}$ exhibits a non-monotonic fluence dependence. For FX--1LO, it increases from $0.190\pm0.002$ eV in pristine ZnO to $0.250\pm0.002$ eV at $1\times10^{15}$ ions/cm$^{2}$, before decreasing to $0.194\pm0.002$ eV at the highest fluence. FX--2LO follows a similar trend, increasing from $0.128\pm0.002$ to $0.234\pm0.002$ eV and subsequently decreasing to $0.141\pm0.002$ eV. These variations substantially exceed the quoted uncertainties and indicate enhanced temperature-dependent energy renormalisation at intermediate fluence within the adopted model. For $D^0X$, the initial increase from $0.106\pm0.002$ to $0.110\pm0.002$ eV is modest, whereas the subsequent decrease to $0.082\pm0.002$ eV is clearly resolved. FX--1LO consistently yields a larger $a_{\mathrm{BE}}$ than the other features, including at $1\times10^{15}$ ions/cm$^{2}$, where its difference from FX--2LO exceeds the stated uncertainties. These parameters characterise the thermal evolution of the individual spectral features and should be interpreted as effective measures of exciton--phonon interaction within the fitting model.

\begin{figure}[!ht]
\centering
\includegraphics[width=1\linewidth]{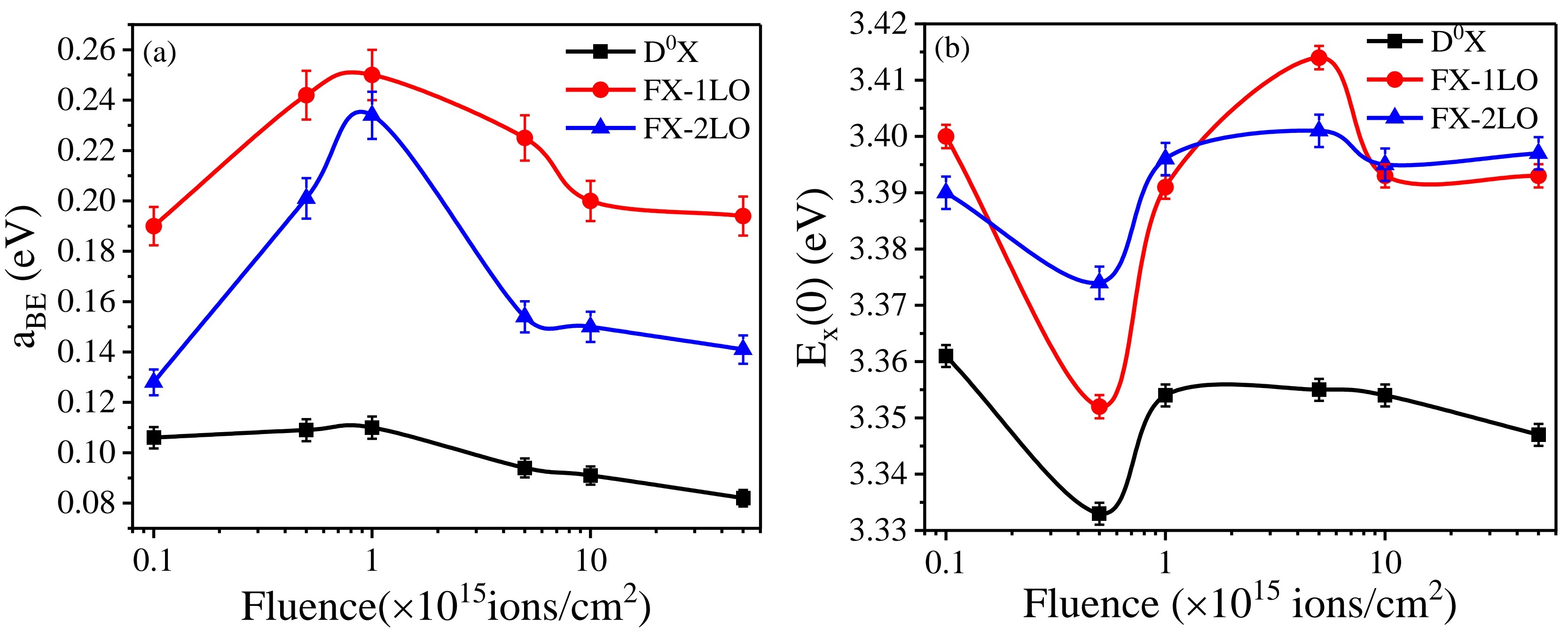}
\caption{Variation of (a) Exciton-Phonon Interaction strength $(a_{\mathrm{BE}})$ and (b) Recombination Energy $(E_x(0))$ with O$^+$ ion fluence for the $D^oX$, FX-1LO and FX-2LO peaks of the NBE emission }
\label{Exciton_Phonon Interaction}
\end{figure}

Figure \ref{Exciton_Phonon Interaction}(a) demonstrates a pronounced fluence dependence of the exciton–phonon interaction strength $a_{\mathrm{BE}}$ for the $D^0X$, FX–1LO, and FX–2LO emission bands. The strength of the interaction increases initially with the fluence of $O^{+}$ and reaches a maximum near $1\times10^{15}$ ions/cm$^{2}$, particularly for free-exciton phonon replicas. The larger $a_{\mathrm{BE}}$ obtained for FX–1LO compared with FX–2LO and $D^0X$ indicates stronger coupling of the free-exciton state to the LO-phonon mode, whereas the progressively lower value for the second-order phonon replica suggests weakening of the effective exciton–phonon coupling for higher-order phonon-assisted transitions. At higher fluences, $a_{\mathrm{BE}}$ decreases, which can be associated with the accumulation of irradiation-induced disorder and the consequent modification of the excitonic and phononic states. Figure \ref{Exciton_Phonon Interaction}(b) shows a distinctly non-monotonic evolution of $E_x(0)$. The initial redshift followed by a partial recovery at intermediate fluence indicates competing effects of irradiation-induced defects, local lattice distortion, and exciton localization. At higher fluences, the recombination energies tend towards saturation, suggesting that the excitonic states approach a highly defect-modified regime. Thus, the interaction strength follows approximately, $a_{\mathrm{BE}}$
$a_{\mathrm{BE}}(FX-1LO)$ $>$ $a_{\mathrm{BE}}(FX-2LO)$ $>$ $a_{\mathrm{BE}}(D^0X)$, particularly around the intermediate fluence.So, the correlated evolution of $a_{\mathrm{BE}}$ and $E_x(0)$ demonstrates that $O^{+}$ irradiation provides an effective means of tailoring both exciton–phonon coupling and excitonic recombination in ZnO.
It should be noted that, the tunable NBE excitonic switching and exciton–phonon coupling provide independent handles for controlling exciton population, recombination, and light–matter interactions, opening opportunities for defect-programmable UV optical switches, modulators, nanophotonic resonators, and phonon-sensitive optoelectronic devices\cite{Jeem2017}. Controlling $a_{\mathrm{BE}}$ through defect concentration and local lattice distortion offers a route to engineer exciton linewidth, energy relaxation, recombination, and phonon-assisted optical transitions.

\section{CONCLUSIONS}

In summary, vertically aligned hexagonal ZnO NRs grown by a hydrothermal process were irradiated with 250 keV $O^{+}$ ions to investigate the formation of defects induced by irradiation and its influence on excitonic and optical properties. The decrease in the oxygen sputtering yield from 1.5 to 0.6 atoms/ion, accompanied by a slight increase in the Zn sputtering yield from 0.3 to 0.5 atoms/ion, indicates preferential oxygen removal and supports the formation of oxygen-vacancy-related defects. Ion irradiation also induces a significant morphological modification, with the initial hexagonal nanorod tips becoming rounded and the diameter of the nanorod decreasing due to sputtering and irradiation-induced thermal effects. The near-band-edge excitonic emission around 380 nm is progressively suppressed and eventually disappears after irradiation, while thermal annealing restores the excitonic peak. This reversible exciton OFF/ON response demonstrates excitation-controlled optical switching associated with irradiation-induced defect states and their thermal recovery. The broad defect-related emission between 400 and 750 nm exhibits superimposed quasi-periodic WGM-like oscillations. Their progressive suppression with increasing fluence is attributed to irradiation-induced changes in NR geometry that modify the optical confinement conditions. Low-temperature PL reveals $D^0X$, FX–1LO, and FX–2LO emissions. Their systematic redshift with temperature is well described by the Bose–Einstein model, confirming the role of the exciton-phonon interaction in band-edge shrinkage. The non-monotonic variation of the extracted $a_{\mathrm{BE}}$ parameter indicates a competing influence of irradiation-induced defects, local lattice distortion, and exciton localization on exciton–phonon coupling. So, $O^{+}$ ion irradiation provides an effective means of tuning defect states, excitonic emission, exciton–phonon interaction, and WGM-like optical resonances in ZnO nanorods, demonstrating their potential for defect-engineered optical switching and nanoscale photonic applications.

\begin{acknowledgments}
The authors acknowledge the financial support from the National Institute of Science Education and Research (NISER) and the Department of Atomic Energy (DAE) under the project grant RIN-4001, SPS. The authors also acknowledge the NISER Ion beam Lab (NIBL) and the staff members Dr. Haraprasanna Lenka and Riratha Chakki, for providing a stable beam for this experiment. 
\end{acknowledgments}

\section*{DECLARATION OF INTERESTS}
The authors declare that they have no known competing financial interests or personal relationships that could have appeared to influence the work reported in this paper.

\section*{Data availability statement}
The data that support the findings of this study are available from the corresponding author upon reasonable request.

\nocite{*}
\bibliography{aipsamp}

\end{document}